\documentclass[prd,onecolumnnopacs,nofootinbib]{revtex4}

\usepackage{amsmath}
\usepackage{graphicx}
\usepackage{amsfonts}
\usepackage{latexsym}
\usepackage{bbold}
\usepackage{wasysym}
\usepackage{calligra}
\usepackage{float}
\usepackage{ulem}
\usepackage{inputenc}
\usepackage{xspace}
\usepackage{url}
\usepackage{epstopdf}
\usepackage{tikz}
\usepackage{cancel} 
\usepackage{amsthm}

\newtheorem*{mydef*}{Definition}

\newcommand{\be}{\begin{equation}}
\newcommand{\ee}{\end{equation}}
\newcommand{\beq} {\begin{equation}}
\newcommand{\eeq} {\end{equation}}
\newcommand{\ba}{\begin{eqnarray}}
\newcommand{\ea}{\end{eqnarray}}

\begin{document}

	%\title{Belifante symmetrization procedure as the flat space limit of Metric-Affine Conservation laws: The cases of QED and QCD}
	\title{Belinfante--Rosenfeld Symmetrization from Metric-Affine Conservation Laws:
Hypermomentum as the Improvement Term---The Cases of QED and QCD}
	
	\author{Damianos Iosifidis}

\affiliation{Scuola Superiore Meridionale, Largo San Marcellino 10, 80138 Napoli, Italy},%Department and Organization
          %  addressline={}, 
           % city={},
          %  postcode={}, 
          %  state={},
           % country={}

\affiliation{INFN– Sezione di Napoli, Via Cintia, 80126 Napoli, Italy}%Department and Organization
          %  addressline={}, 
           % city={},
          %  postcode={}, 
          %  state={},
           % country={}

	\email{d.iosifidis@ssmeridionale.it}
	
	\date{\today}
	\begin{abstract}
		
		We derive the Belinfante--Rosenfeld symmetrization procedure from the metric-affine conservation law by means of an affine lift of flat-spacetime field theories. Following minimal coupling to an independent affine connection, variation of the action with respect to the connection defines the hypermomentum current. Taking the flat-spacetime limit of the resulting conservation law, we recover the Belinfante--Rosenfeld relation, with the divergence of the hypermomentum current reproducing the Belinfante improvement term. We thoroughly study the form of the couplings with this property and their physical significance. The construction is illustrated for Quantum Electrodynamics and Quantum Chromodynamics.
		
	\end{abstract}
	
	\maketitle
	
	\allowdisplaybreaks
	
	%\newpage
	
	%\tableofcontents
	%\newpage

\section{Introduction}
The energy-momentum tensor is one of the fundamental objects in field theory, encoding the distribution and flow of energy and momentum. In flat Minkowski spacetime, applying Noether's theorem to spacetime translations leads to the canonical energy-momentum tensor, whose conservation reflects translational invariance \cite{Noether1918,Weinberg1995}. However, the canonical tensor is generally not symmetric, and much worse not even gauge invariant, particularly in theories containing fields with intrinsic spin, and therefore does not possess all the properties required for a physically satisfactory stress-energy tensor in Minkowski space \cite{Weinberg1995,PeskinSchroeder1995}. The Belinfante--Rosenfeld procedure remedies this by adding an identically conserved improvement term constructed from the spin current, yielding a symmetric energy-momentum tensor while leaving the conserved four-momentum, along with the conservation law, unchanged \cite{Belinfante1940,Belinfante1940b,Rosenfeld1940,Hehl1976EnergyTensor}. For modern treatments of the topic, following the results of \cite{Hehl1976EnergyTensor}, see for instance \cite{GotayMarsden1992,Forger2004Currents,Julia:1998ys,GamboaSaravi2004,Pons2009Noether,Blaschke2016}. Although highly successful and widely adopted, this procedure is often regarded as somewhat \textit{ad hoc}, since the improvement is introduced by hand rather than arising uniquely and precisely from the underlying variational principle. This has motivated the search for alternative formulations of the energy-momentum tensor, particularly in gauge theories and gravitational settings, where one seeks constructions that follow more directly from the fundamental symmetries or geometric structure of the theory. For instance one way to derive the symmetric (Hilbert) energy-momentum tensor is by coupling matter to the metric, deriving the response under metric variations (i.e. the Hilbert variation) and consequently taking the flat space Minkowski limit. Such procedure, however, is somewhat limited since it breaks down in certain cases, like the Dirac field for instance where the introduction of a vielbein is unavoidable in order to couple fermions to gravity.
In this regard, a universal procedure, motivated also by geometric considerations, is mostly welcomed.  More precisely, understanding the Belinfante improvement from a geometric perspective provides a deeper fundamental interpretation of the symmetrization procedure, clarifying also the role of spin, local symmetries, and conservation laws within a unified framework.

A particularly natural setting for such an investigation is metric-affine geometry, in which the metric and affine connection are treated as independent geometric variables \cite{Hehl1976,Hehl1995}. Unlike Riemannian geometry, the metric-affine framework admits both torsion and nonmetricity, allowing intrinsic spin and other matter currents, such as dilation and shear, to couple directly to spacetime geometry. Within this broader geometric setting, the various conserved currents associated with matter fields---including energy-momentum, spin, and hypermomentum---arise on an equal footing through variational principles and the associated Noether identities \cite{Hehl1995,HehlObukhov2003}. 
%This suggests that the Belinfante symmetrization may admit a genuine geometric interpretation, rather than appearing as an external improvement of the canonical tensor. From this perspective, the Belinfante procedure can be viewed as a manifestation of deeper geometric identities relating the matter currents to the underlying spacetime structure, offering new insight into the interplay between symmetry, conservation laws, and gravitation \cite{ObukhovRubilar2006,ObukhovHehl2014}.
From this perspective, the Belinfante procedure is not merely an algebraic improvement of the canonical energy-momentum tensor, but rather a consequence of the underlying geometric structure of spacetime and the Noether identities associated with local gauge symmetries. This viewpoint places the symmetrization procedure on a more fundamental level and naturally unifies the concepts of canonical energy-momentum, Hilbert energy-momentum  and hypermomentum tensor within this extended framework of metric-affine geometry \cite{Hehl1995,HehlObukhov2003,ObukhovRubilar2006}. In particular we will show how hypermomentum is the improvement term that comes as a left-over piece of the affine geometry, connection-matter couplings, in the flat space limit.

The paper is organized as follows.  Firstly we communicate the basic geometric ingredients of Metric-Affine Geometry and consequently introduce the energy tensors of MAG and the associated generalized conservation laws, or better stated, the 'balance equations' in this setting.
Then, we consider matter fields minimally coupled to a general affine connection within the framework of this extended geometry. Varying the matter action with respect to the  affine connection, we obtain the corresponding hypermomentum current, and consequently take the flat space limit by trivializing the connection and setting the metric to the Minkowski one (i.e. $\Gamma^{\lambda}{}_{\mu\nu}\equiv 0$, $g_{\mu\nu}\equiv\eta_{\mu\nu}=(-1,1,1,1)$). We then investigate the flat-spacetime limit of the Metric-Affine conservation laws and show explicitly that the Belinfante--Rosenfeld relation is nothing else than the Minkowski limit of these balance equations. In particular, the divergence of the antisymmetric part of the hypermomentum current reproduces the Belinfante improvement term, thereby identifying the latter as the flat-spacetime remnant of the more general geometric structure encoded in metric-affine gravity. This establishes a direct geometric origin for the Belinfante symmetrization procedure and clarifies the role of hypermomentum as the quantity underlying the improvement of the canonical energy-momentum tensor. The examples we investigate are those of QED and QCD. By performing an \textit{'affine lift'}, we explicitly show that this geometric approach gives rise to the correct symmetric energy-momentum tensors. We discuss the general form of such improvement couplings and under what conditions the hypermomentum qualifies as an improvement term. 
Finally, we provide a step by step algorithm for the simple application of our procedure and conclude our results.

\section{The geometric setup}

We  consider a generic n-dimensional manifold equipped with a metric $g_{\mu\nu}$ and a general linear connection $\Gamma^{\lambda}{}_{\mu\nu}$. In this metric-affine space we  define the torsion, curvature and nonmetricity tensors according to
\begin{subequations}
\begin{align}
S_{\mu\nu}^{\;\;\;\lambda}&:=\Gamma^{\lambda}_{\;\;\;[\mu\nu]}
\;\;, \;\; \\
R^{\mu}_{\;\;\;\nu\alpha\beta}&:= 2\partial_{[\alpha}\Gamma^{\mu}_{\;\;\;|\nu|\beta]}+2\Gamma^{\mu}_{\;\;\;\rho[\alpha}\Gamma^{\rho}_{\;\;\;|\nu|\beta]} \;\;,\label{R} 
  \\
Q_{\alpha\mu\nu}&:=- \nabla_{\alpha}g_{\mu\nu}
\end{align}
\end{subequations}
respectively. By taking contractions one obtains one torsion and two nonmetricity vectors, repsectively defined by
\beq
S_{\mu}:=S_{\mu\lambda}^{\;\;\;\;\lambda} \;\;, \;\;\;
Q_{\alpha}:=Q_{\alpha\mu\nu}g^{\mu\nu}\;,\;\; q_{\nu}=Q_{\alpha\mu\nu}g^{\alpha\mu}
\eeq
The deviation from the Riemannian geometry is quantified by the so-called distortion tensor \cite{hehl1995metric}
\begin{gather}
N^{\lambda}_{\;\;\;\;\mu\nu}:=\Gamma^{\lambda}_{\;\;\;\mu\nu}-\widetilde{\Gamma}^{\lambda}_{\;\;\;\mu\nu}=
\frac{1}{2}g^{\alpha\lambda}(Q_{\mu\nu\alpha}+Q_{\nu\alpha\mu}-Q_{\alpha\mu\nu}) -g^{\alpha\lambda}(S_{\alpha\mu\nu}+S_{\alpha\nu\mu}-S_{\mu\nu\alpha}) \label{N}
\end{gather}
Accordingly, torsion and nonmetricity are derived from the distortion through the relations 
\cite{hehl1995metric,iosifidis2019metric}
\beq
S_{\mu\nu\alpha}=N_{\alpha[\mu\nu]}\;\;,\;\;\; Q_{\nu\alpha\mu}=2 N_{(\alpha\mu)\nu} \label{QNSN}
\eeq
which can be readily verified by the above definitions.

\begin{mydef*}{\underline{Affine Lift}.}
We define the affine-lift as the map
\[
(\mathbb{R}^{1,n-1},\eta)
\mapsto
(M,g,\Gamma),
\]
sending Minkowski spacetime to a generic metric-affine space, with arbitrary linear connection and metric.
\end{mydef*}
The above definition will be essential for the symmetrization procedure we are going to use throughout this work. 
This is the geometric minimum we shall need for the rest of our discussion.

\section{Sources and Conservation laws of MAG}

Firstly, let us recall the sources of Metric-Affine Gravity. These are the  Canonical and Metrical Energy-Momentum Tensors, that are respectively given by \cite{hehl1995metric}\footnote{Here $e_{\mu}^{\;\;a}$ is as usual the vielbein which defines the metric through $g_{\mu\nu}=e_{\mu}^{\;\;a}e_{\nu}^{\;\;a}g_{ab}$ where $g_{ab}$ is the tangent space metric. If the vielbein is taken to be orthonormal, the latter becomes the Minkowski metric $\eta_{ab}$.}
\beq
t^{\mu}_{\;\;\nu}:=-\frac{\partial \mathcal{L}_{M}}{\partial(\nabla_{\mu}\psi^{A})}\nabla_{\nu}\psi^{A}+\delta_{\nu}^{\mu}\mathcal{L}_{M}=\frac{1}{\sqrt{-g}} \frac{\delta (\sqrt{-g}\mathcal{L}_{\text{M}})}{\delta {e_\mu}^c} e_{\nu}{}^{c} \,.
\eeq
and
\beq
T_{\mu \nu} := - \frac{2}{\sqrt{-g}} \frac{\delta S_{\text{M}}}{\delta g^{\mu \nu}} = - \frac{2}{\sqrt{-g}} \frac{\delta (\sqrt{-g} \mathcal{L}_{\text{M}})}{\delta g^{\mu \nu}} \,, \label{metrical}
\eeq
 along with the (intrinsic) Hypermomentum tensor  \cite{hehl1976hypermomentum,hehl1978hypermomentum}
 which is defined as the variation of the matter part of the action with respect to the connection, namely\footnote{It is important to point out that  hypermomentum has both an orbital as well as an intrinsic part, see \cite{Gronwald:1991st}.}
\beq
{\Delta_\lambda}^{\mu \nu} := - \frac{2}{\sqrt{-g}} \frac{\delta S_{\text{M}}}{\delta {\Gamma^\lambda}_{\mu \nu}} = - \frac{2}{\sqrt{-g}} \frac{\delta (\sqrt{-g} \mathcal{L}_{\text{M}})}{\delta {\Gamma^\lambda}_{\mu \nu}} \,.
\eeq
The latter tensor is of ultimate importance since it encapsulates  the micro-properties of matter (spin, dilation and shear). It can be split into these three irreducible pieces of spin, dilation and shear according to (see \cite{hehl1995metric})
\beq
\Delta_{\alpha\mu\nu}=\Sigma_{\alpha\mu\nu}+\frac{1}{n}g_{\alpha\mu}\Delta_{\nu}+\hat{\Delta}_{\alpha\mu\nu} \label{hypsplit}
\eeq
with
\beq
\sigma^{\mu\nu\alpha}:=\Delta^{[\mu\nu]\alpha} \;\;\; (Spin) \label{spin}
\eeq
\beq
\Delta^{\nu}:=\Delta^{\alpha\mu\nu}g_{\alpha\mu} \;\;\; (Intrinsic \;Dilation)
\eeq
\beq
\Sigma^{\mu\nu\alpha}:=\Delta^{(\mu\nu)\alpha}-\frac{1}{n}g^{\mu\nu}\Delta^{\alpha} \;\;\; (Intrinsic\; Shear)
\eeq
The introduction of the full Hypermomentum can be understood as follows:
With the conventional energy-momentum $T^{\mu\nu}$ alone, we describe basically MONOpole type
of matter, if we introduce spin, we take care of the DIpolar structure
of matter and finally, if we introduce (dilation and shear) hypermomentum, the
QUADRUpolar structure of matter, if present, is taken care of, see Hehl and Ne'eman (1990)\cite{Hehl:1990SpacetimeMicrostructure}.

	Working in the exterior calculus framework, from the GL and Diffeomorphism invariance of the matter action one gets the conservation laws \cite{hehl1995metric,Iosifidis:2020gth,obukhov2013conservation} (expressed in a holonomic frame):

	\beq
	t^{\mu}_{\;\;\lambda}
	= T^{\mu}_{\;\;\lambda}-\frac{1}{2 \sqrt{-g}}(2S_{\nu}-\nabla_{\nu})(\sqrt{-g}\Delta_{\lambda}^{\;\;\mu\nu}) \label{cc1}
	\eeq
	\beq
	\frac{1}{\sqrt{-g}}(2S_{\mu}-\nabla_{\mu})(\sqrt{-g}t^{\mu}_{\;\;\alpha})=-\frac{1}{2} \Delta^{\lambda\mu\nu}R_{\lambda\mu\nu\alpha}+\frac{1}{2}Q_{\alpha\mu\nu}T^{\mu\nu}+2 S_{\alpha\mu\nu}t^{\mu\nu} \label{cc2}
	\eeq
This is the most general form of the conservation laws or more appropriately called the 'balance equations' of Metric-Affine Gravity. In (\ref{cc1}) one immediately recognizes a generalization of the Belinfante symmetrization procedure \cite{hehl1995metric}. 

\subsection{Flat Space Limit}
Now an idea comes, what if we take the flat Minkowski space limit of the previous relativistic Metric-Affine conservation laws? This is obtained by switching off gravity, namely we set $g_{\mu\nu}\rightarrow \eta_{\mu\nu}$ and $\Gamma^{\lambda}{}_{\mu\nu}\rightarrow 0$ which also imply $S_{\mu\nu}{}{}^{\lambda}\equiv 0$, $R^{\lambda}{}_{\mu\nu\alpha}\equiv 0$ and $Q_{\alpha\mu\nu}\equiv 0$. Accordingly, all covariant derivatives reduce to the partial $\partial_{\mu}$ ones and the  conservation laws of MAG (\ref{cc2}) and (\ref{cc1}) take the neat form\footnote{The same equations were derived already in \cite{HEHL1977432} almost 50 years ago, by a relativistic extension of the
group $SL(3, R)$ to the general affine group
$GA(4, R)$.}
\beq
\partial_{\mu}t^{\mu\nu}=0 \label{cl1}
\eeq
\beq
t^{\mu\nu}=T^{\mu\nu}+\frac{1} {2}\partial_{\alpha}\Delta^{\nu\mu\alpha} \label{cl2}
\eeq
The first conclusion by looking the latter equations, is that the canonical energy-momentum tensor is in general the one that is properly conserved. It is therefore a physical one that describes the proper energy content of matter fields. From the second conservation law we see that the metrical energy-momentum tensor is a nicely constructed byproduct; it is derived once the canonical and the hypermomentum currents are known. It is also clear that the metrical energy-momentum tensor $T^{\mu\nu}$ is generally non-conserved but only when the hypermomentum has the off-shell property
\beq
\partial_{\mu}\partial_{\nu}\Delta_{\lambda}{}^{\mu\nu}\equiv 0
\eeq
This is certainly true whenever\footnote{This should not be confused with the spin part (\ref{spin}) which is antisymmetric in the first pair of indices. It should be noted also that this solution is capable but not strictly necessary.}
\beq
\Delta_{\nu\mu\alpha}=\Delta_{\nu[\mu\alpha]} \label{improvedhyper}
\eeq
If this is the case, then the hypermomentum acts as an improvement term rendering the energy-momentum tensor symmetric and conserved through the re-writing
\beq
T^{\mu\nu}=t^{\mu\nu}-\frac{1} {2}\partial_{\alpha}\Delta^{\nu\mu\alpha} \;\;, \;\; \partial_{\mu}T^{\mu\nu}=0 \label{flatCL}
\eeq
Now another idea comes, what if we consider the minimal substitution $\partial_{\mu}\rightarrow \nabla_{\mu}$ in the initial matter action that we are interested in obtaining the symmetric and conserved energy-momentum tensor, compute the associated hypermomentum and then substitute it in (\ref{flatCL})? If the hypermomentum has the antisymmetry property (\ref{improvedhyper}), then the resulting energy-momentum tensor $T^{\mu\nu}$ will also be conserved apart from symmetric. This   idea came to me as an insight four years ago in the summer of 2022.\footnote{This happened while going through the  'The Classical Theory of Fields' \cite{landau1975classical} by Landau \& Lifshitz. Perhaps this proves that going through the classics is always a good idea to generate new constructive thoughts and novel ideas.  }  I immediately applied it to the Maxwell field and quite miraculously it worked; the extra term I obtained from hypermomentum was exactly the one derived from the Belinfante procedure.
Then later with  colleagues, we considered an extended variety of examples like the Dirac field, Abelian gauge fields, the conformably coupled scalar field as well as some non-unitary scalar field theories and showed that indeed in all cases the hypermomentum  of this type acted as the improvement term \cite{Iosifidis:2025sjx}. All of these applications revolved around single non-interacting fields.\footnote{The term non-interacting here means interaction with other fields apart from geometry.} We shall now apply it to the physically interesting and realistic cases  of Quantum Electrodynamics (QED) and Quantum Chromodynamics (QCD).

\section{The energy-momentum tensors of QED and QCD derived by geometric considerations}

\subsection{QED Energy-momentum tensor }

We shall start with QED Lagrangian in flat Minkowski spacetime,  namely 
\beq
\mathcal{L}_{\text{QED}} =- \frac{i}{2}\Big( \bar{\psi} \overrightarrow{\cancel{D}} \psi-\bar{\psi}\overleftarrow{\cancel{D}}\psi \Big)-i m\bar{\psi}\psi - \frac{1}{4} F_{\mu\nu} F^{\mu\nu} \label{LQED}
\eeq
where $\cancel{D}:=\gamma^{\mu}D_{\mu}$ and  with $D_{\mu}$ being the gauge covariant derivative acting on the spinor field and its adjoint as
\beq
\overrightarrow{D}_{\mu}\psi:=\partial_{\mu}\psi-ie A_{\mu}\psi
\eeq
and 
\beq
\bar{\psi}\overleftarrow{D}_{\mu}: = \partial_{\mu} \bar{\psi} + ie A_{\mu} \bar{\psi}
\eeq
respectively. 
As usual $F_{\mu\nu}:=\partial_{\mu}A_{\nu}-\partial_{\nu}A_{\mu}$ is the field strength tensor associated to the gauge field $A_{\mu}$. The equations of motion for the above QED Lagrangian read
\beq
\partial_{\mu}F^{\mu\nu}-e \bar{\psi}\gamma^{\nu}\psi =0\label{feA}
\eeq
\beq
\overrightarrow{\cancel{D}}\psi=-m\psi
\eeq
\beq
\bar{\psi}\overleftarrow{\cancel{D}}=m\bar{\psi} \label{fepsi2}
\eeq
and the associated canonical energy-momentum tensor is readily derived to be
\beq
t^{\mu\nu}=F^{\mu\lambda}\partial^{\nu}A_{\lambda}-\frac{1}{4}\eta^{\mu\nu} F^{\alpha\beta}F_{\alpha\beta}+\frac{i}{2}\Big( \bar{\psi}\gamma^{\mu}\partial^{\nu}\psi-(\partial^{\nu}\bar{\psi})\gamma^{\mu}\psi \Big) \label{canonicalQEC}
\eeq
which is obviously neither symmetric nor gauge invariant. It is nevertheless, conserved i.e.  $\partial_{\mu}t^{\mu}{}_{\nu}=0$ when the equations of motion for the fields hold true. We shall show how one can obtain the symmetric (metrical) energy-momentum tensor by uplifting the Minkowski space to the full Metric-Affine one\footnote{To be more precise to the Riemman-Cartan space since we assume vanishing non-metricity in order to maintain finite dimensional representations for the spinor fields.}, compute the associated hypermomentum and then go back to the initial flat Minkowski space. This is the essense of what we shall call an 'affine lift'. The first step is to apply a minimal substitution to both the field strength as well as the spinor field derivatives, namely we simultaneously promote
\beq
F_{\mu\nu}:=\partial_{\mu} A_{\nu}-\partial_{\nu}A_{\mu} \mapsto \boxed{\bold{F}_{\mu\nu}:=\nabla_\mu A_\nu-\nabla_\nu A_\mu=F_{\mu\nu}+2 S_{\mu\nu}{}{}^{\lambda}A_{\lambda}}
\eeq
and
\beq
\overrightarrow{D}_{\mu}\psi:=\partial_{\mu}\psi-ie A_{\mu}\psi \mapsto \boxed{\overrightarrow{\bold{D}}_{\mu}\psi:=\partial_{\mu}\psi-ie A_{\mu}\psi +\frac12\omega_{ab \mu}\Sigma^{ab}\psi}
\eeq
and 
\beq
 \bar{\psi}\overleftarrow{D}_{\mu}: = \partial_{\mu} \bar{\psi} + ie A_{\mu} \bar{\psi}  \mapsto \boxed{\bar{\psi}\overleftarrow{\bold{D}}_{\mu}: = \partial_{\mu} \bar{\psi} + ie A_{\mu} \bar{\psi}-\frac12\omega_{ab\mu}\bar\psi\Sigma^{ab}}
\eeq
%Now moving on to the divergence term $\partial_{\alpha}A^{\alpha}$ one must be careful in recalling that there is one and only one way to generalize the divergence operator to the curved case and this makes no reference to a connection. Indeed, the proper definition of the divergence of the vector field $A^{\mu}$ to curved spacetime reads
%\beq
%\partial_{\alpha}A^{\alpha}\mapsto \boxed{\frac{1}{\sqrt{-g}}\partial_{\alpha}(\sqrt{-g}A^{\alpha})}
%\eeq
%This form then ensured that the Gauss law can be applied to the integral $\int d^{4}x \partial_{\alpha}(\sqrt{-g}A^{\alpha})$. Of course one may re-write the above expression in a covariant manner using a connection, but the point is the divergence depends only on the metric and not on the connection. Using either a general affine connection $\nabla_{\alpha}  $ or the Levi-Civita one $\widetilde{\nabla}_{\alpha}$, one can identically write
%\beq
%\frac{1}{\sqrt{-g}}\partial_{\alpha}(\sqrt{-g}A^{\alpha})=\frac{1}{\sqrt{-g}}(\nabla_{\alpha}-2 S_{\alpha})(\sqrt{-g}A^{\alpha})=\widetilde{\nabla}_{\alpha}A^{\alpha}
%\eeq
Under the above replacements, the affinely-lifted QED Lagrangian is obtained as
\beq
\bold{\mathcal{L}^{aff}_{\text{QED}}} =- \frac{i}{2}\Big( \bar{\psi} \overrightarrow{\cancel{\bold{D}}} \psi-\bar{\psi}\overleftarrow{\cancel{\bold{D}}}\psi \Big)- i m\bar{\psi}\psi - \frac{1}{4} \bold{F}_{\mu\nu} \bold{F}^{\mu\nu}  \label{LaffQED}
\eeq
when now the gamma matrices are defined as $\gamma^{\mu}:=e^{\mu}{}_{a}\gamma^{a}$ where  $e_{\mu}{}^{a}$ is the vielbein or tetrad, $a$ denotes tangent space index and $\gamma^{a}$ are the usual constant element gamma matrices. Now the procedure goes as follows. From the uplifted Lagrangian (\ref{LaffQED}) we compute the variational derivative with respect to the affine connection, namely the hypermomentum and consequently go back to flat space by trivializing the connection and then setting the metric to be the Minkowski, according to the scheme:
\beq
\Delta_{\lambda}{}^{\mu\nu}:=-\frac{2}{\sqrt{-g}}\frac{\delta (\sqrt{-g}\bold{\mathcal{L}^{aff}_{\text{QED}}})}{\delta \Gamma^{\lambda}{}_{\mu\nu}}\Big|_{\Gamma=0,\; g=\eta}
\eeq
We explicitly derive
\beq
\Delta_{\lambda}{}^{\mu\nu}=2 A_{\lambda}F^{\mu\nu}-\frac{i}{2}\bar{\psi}\Big( \gamma_{\nu}\Sigma^{\mu}{}_{\lambda}+\Sigma^{\mu}{}_{\lambda}\gamma^{\nu}\Big) \psi
\eeq
as the leftover piece in the flat space limit. In the above. we have defined, as usual, $4\Sigma^{\mu\nu}=[\gamma^{\mu},\gamma^{\nu}]$. Then, using the field equations (\ref{feA})-(\ref{fepsi2}) we compute the on-shell derivative\footnote{We also use the well known identity $[\gamma^{\lambda},\Sigma^{\mu\nu}]=\eta^{\mu\lambda}\gamma^{\nu}-\eta^{\nu\lambda}\gamma^{\mu}$.}
\beq
\frac{1}{2}\partial_{\lambda}\Delta^{\nu\mu\lambda}=(\partial_{\lambda}A^{\nu})F^{\mu\lambda}-e \bar{\psi}A^{\nu}\gamma^{\mu}\psi-\frac{i}{4}\left( \bar{\psi}\gamma^{\nu}\overrightarrow{D}^{\mu}\psi -\bar{\psi}\gamma^{\mu}\overrightarrow{D}^{\nu}\psi -(\bar{\psi}\overleftarrow{D}^{\mu})\gamma^{\nu}\psi+(\bar{\psi}\overleftarrow{D}^{\nu})\gamma^{\mu}\psi \right)
\eeq
We substitute the latter along with (\ref{canonicalQEC}) into the master relation (\ref{cl2}) and lo and behold
\begin{gather}
T^{\mu\nu}=F^{\mu\lambda}F^{\nu}{}_{\lambda}+\eta^{\mu\nu}\mathcal{L}_{QED}+\frac{i}{4}\Big( \bar{\psi}\gamma^{\mu} \overrightarrow{D}^{\nu}\psi +\bar{\psi}\gamma^{\nu}\overrightarrow{D}^{\mu}\psi \Big)-\frac{i}{4}\Big(  (\bar{\psi}\overleftarrow{D}^{\nu})\gamma^{\mu}\psi+(\bar{\psi}\overleftarrow{D}^{\mu})\gamma^{\nu}\psi\Big) 
\end{gather}
we arrive to the symmetrized energy-momentum tensor of QED! In the above we have also used the fact that on-shell $\mathcal{L}_{D}=0$, where $\mathcal{L}_{D}$ is the Dirac contribution in (\ref{LQED}). Our result for the symmetrized energy-momentum tensor is in perfect agreement with the standard result \cite{Embacher1986,Montesinos:2006th} but we arrived at it by purely geometric considerations and a much less painful, almost trivial, computation.

\subsection{The case of QCD}

The construction presented above extends naturally to non-Abelian gauge theories. As an illustrative example we consider Quantum Chromodynamics (QCD), whose flat-spacetime Lagrangian is given by
\beq
\mathcal{L}_{\rm QCD}
=
-\frac14 G_{\mu\nu}^{\,a}G^{a\mu\nu}
-
\sum_{f}
\left[
\frac{i}{2}
\Big(
\bar\psi_f\overrightarrow{\cancel D}\psi_f
-
\bar\psi_f\overleftarrow{\cancel D}\psi_f
\Big)
+im_f\bar\psi_f\psi_f
\right]
\eeq
where
\beq
\overrightarrow{D}_{\mu}\psi_f
=
\partial_{\mu}\psi_f
-ig_sA_{\mu}^{\,a}T^{a}\psi_f,
\qquad
\bar\psi_f\overleftarrow{D}_{\mu}
=
\partial_{\mu}\bar\psi_f
+ig_sA_{\mu}^{\,a}\bar\psi_fT^{a},
\eeq
and the Yang--Mills field strength is, as usual,
\beq
G_{\mu\nu}^{\,a}
=
\partial_{\mu}A_{\nu}^{\,a}
-
\partial_{\nu}A_{\mu}^{\,a}
+
g_sf^{abc}A_{\mu}^{\,b}A_{\nu}^{\,c}
\eeq
Here $A_{\mu}^{a}$ are the gluon gauge fields, $T^a$ are the generators of the Lie algebra $\mathfrak{su}(3)$ in the fundamental representation of the gauge group $SU(3)$, $f^{abc}$ denote the corresponding  totally antisymmetric structure constants, and $f=1,2,...,6$ labels the quark flavors.
The  field equations of the above  QCD Lagrangian read
\beq
(D_{\mu}G^{\mu\nu})^{a}
=
g_s
\sum_{f}
\bar\psi_f\gamma^{\nu}T^{a}\psi_f
\eeq
together with
\beq
\overrightarrow{\cancel D}\psi_f=-m_f\psi_f
\qquad
\bar\psi_f\overleftarrow{\cancel D}=m_f\bar\psi_f 
\eeq
The associated canonical energy--momentum tensor is straightforwardly computed to be
\beq
t^{\mu\nu}
=
G^{a\mu\lambda}\partial^{\nu}A_{\lambda}^{\,a}
-\frac14\eta^{\mu\nu}
G_{\alpha\beta}^{\,a}G^{a\alpha\beta}
-
\sum_{f}
\frac{i}{2}
\Big(
\bar\psi_f\gamma^{\mu}\partial^{\nu}\psi_f
-
(\partial^{\nu}\bar\psi_f)\gamma^{\mu}\psi_f
\Big)
\eeq
which is conserved on shell but is neither symmetric nor gauge invariant. Note that it can be made gauge invariant by a so-called pseudo-gauge transformation that was introduced by Hehl \cite{Hehl1976}.\footnote{Such a transformation has non-trivial consequences in spin hydrodynamics \cite{Becattini:2018duy,Becattini:2025oyi,Montenegro:2026phf,Zhang:2026DilationSpin}.}

Proceeding exactly as in the Abelian case, we now promote
\beq
G_{\mu\nu}^{\,a}
\longmapsto
\boxed{\mathbf G_{\mu\nu}^{\,a}
:=
\nabla_{\mu}A_{\nu}^{\,a}
-
\nabla_{\nu}A_{\mu}^{\,a}
+
g_sf^{abc}A_{\mu}^{\,b}A_{\nu}^{\,c}
=
G_{\mu\nu}^{\,a}
+
2S_{\mu\nu}{}^{\lambda}A_{\lambda}^{\,a}}
\eeq
while simultaneously replacing
\beq
\overrightarrow{D}_{\mu}\psi_{f} \mapsto \boxed{\overrightarrow{\bold{D}}_{\mu}\psi_{f}:=\partial_{\mu}\psi_{f}-i g_{s} A_{\mu}^{a}T^{a}\psi_{f}  +\frac12\omega_{ab \mu}\Sigma^{ab}\psi_{f}}
\eeq
for every quark flavor, and similarly for the adjoint. Since the affine connection acts only on spacetime indices, the non-Abelian self-interaction term remains unaffected by the affine lift. Consequently, the affinely uplifted QCD Lagrangian is constructed as
\beq
\bold{\mathcal{L}_{QCD}^{aff}}
=
-\frac14 \bold{G}_{\mu\nu}^{\,a}\bold{G}^{a\mu\nu}
-
\sum_{f}
\left[
\frac{i}{2}
\Big(
\bar{\psi} \overrightarrow{\cancel{\bold{D}}} \psi-\bar{\psi}\overleftarrow{\cancel{\bold{D}}}\psi \Big)+ i m\bar{\psi}\psi
\right]
\eeq

Variation of the affinely uplifted action with respect to the independent affine connection  and subsequent  reduction  to Minkowski spacetime yield the hypermomentum current
\beq
\Delta_{\lambda}{}^{\mu\nu}
=
2A_{\lambda}^{\,a}G^{a\mu\nu}
-
\frac{i}{2}
\sum_{f}
\bar\psi_f
\left(
\gamma^{\nu}\Sigma^{\mu}{}_{\lambda}
+
\Sigma^{\mu}{}_{\lambda}\gamma^{\nu}
\right)
\psi_f 
\eeq
Taking the divergence of the above expression and employing the Yang--Mills and Dirac equations, one finds that
$\frac12\partial_{\lambda}\Delta^{\nu\mu\lambda}$
reproduces precisely the Belinfante improvement term as in QED, and consequently recovers the symmetric and \underline{conserved} energy--momentum tensor
\beq
T^{\mu\nu}
=
G^{a\mu\lambda}G^{a\nu}{}_{\lambda}
+\eta^{\mu\nu}\mathcal{L}_{\rm QCD}
+
\frac{i}{4}
\sum_{f}
\left[
\bar\psi_f\gamma^{\mu}\overrightarrow{D}^{\,\nu}\psi_f
+\bar\psi_f\gamma^{\nu}\overrightarrow{D}^{\,\mu}\psi_f
-(\bar\psi_f\overleftarrow{D}^{\,\nu})\gamma^{\mu}\psi_f
-(\bar\psi_f\overleftarrow{D}^{\,\mu})\gamma^{\nu}\psi_f
\right]
\eeq
where the Dirac contribution vanishes on shell for each flavor. The result is in perfect agreement with \cite{Nielsen1977}. Thus, as in QED, the Belinfante--Rosenfeld tensor follows directly from the metric-affine conservation law, demonstrating that the geometric origin of the Belinfante improvement is independent of the internal gauge group and remains valid for the full non-Abelian gauge theory.

Our result is also in perfect agreement with previous works and approaches on the matter \cite{KugoOjima1979,Ji:1996ek,Leader:2013jra,Zoller:2014dca,Freese:2025glz,Lorce:2021xku}. However, there are at least three key points, that, in our opinion, make our formulation here superior. Most importantly, the calculations performed here to arrive to the final result were quite elegant and simple, whereas in these other works one arrives at the same result after some quite elaborate, many page calculations. Secondly, in our formulation we have a precise step by step guideline to follow without any added ad hoc considerations. And thirdly, in our formulation the improvement term has a precise geometric meaning as the left-over hypermomentum produced by the affine uplift. The very fact that the prescription works for all physically relevant cases, strengthens even more the validity of this approach.

\section{On the physical content of hypermomentum as an improvement term}

\subsection{When does Hypermomentum qualify as an improvement term?}

As seen from (\ref{cl2}) the hypermomentum acts as an improvement term, namely it symmetrizes and most importantly it does not spoil the conservation law for the energy momentum tensor iff the double derivative on it vanishes of shell, i.e.
\beq
\partial_{\mu}\partial_{\nu}\Delta_{\lambda}{}^{\mu\nu}\equiv 0
\eeq
This condition is true whenever hypermomentum is antisymmetric in the last pair of indices ($\Delta_{\lambda}{}^{\mu\nu}=\Delta_{\lambda}{}^{[\mu\nu]}$). Of course, this condition is capable but not strictly necessary. Note that this antisymmetry is always guaranteed when the  matter-connection couplings depend explicitly only on torsion, but this is not the only possibility. Interestingly enough, hypermomentum can act as an improvement term even in the presence of nonmetricity, given that only certain parts of nonmetricity  couple to matter. To see this, start with the definition of hypermomentum and apply the chain rule using also the definitions of torsion and nonmetricity. We readily obtain
\beq
\Delta_{\lambda}{}^{\mu\nu}= W_{\lambda}{}^{\mu\nu}+C^{\nu\mu}{}{}_{\lambda}+C^{\nu\;\;\;\mu}_{\;\;\lambda}=W_{\lambda}{}^{\mu\nu}+2 C^{\nu\mu}{}{}_{\lambda}\label{DeltaWC}
\eeq
where  
\beq
W_{\lambda}{}^{\mu\nu}=W_{\lambda}{}^{[\mu\nu]}:=-\frac{2}{\sqrt{-g}}\frac{\delta(\sqrt{-g}\mathcal{L}_{M})}{\delta S_{\mu\nu}{}^{\lambda}}\;\;  ,\;\;C^{\nu}{}_{\alpha\beta}=C^{\nu}{}_{(\alpha\beta)}:=-\frac{2}{\sqrt{-g}}\frac{\delta(\sqrt{-g}\mathcal{L}_{M})}{\delta Q_{\nu}{}^{\alpha\beta}}
\eeq
The first term in (\ref{DeltaWC}) is antisymmetric in $\mu,\nu$ by default which proves our previous claim that torsion couplings act as improvement terms. As for the second piece, we see that in general is asymmetric in $\mu,\nu$ and therefore generally spoils the improvement character.

 Nevertheless, if only the  antisymmetric part $Q_{[\mu\nu]\alpha}$ of nonmetricity is involved then the second term is also an improvement term since the tensor 
\beq
C^{\nu\mu}{}{}_{\lambda}:=-\frac{2}{\sqrt{-g}}\frac{\delta(\sqrt{-g}\mathcal{L}_{M})}{\delta Q_{[\nu\mu]}{}{}^{\lambda}} 
\eeq
is by construction antisymmetric in $\mu,\nu$. With this observation we can state the form a fairly general matter MAG Lagrangian for which the associated hypermomentum is always an improvement term a la Belinfante. Hypermomentum is an improvement term\footnote{It is important to keep in mind that we are not claiming that this is the most general Lagrangian, it is rather a subset of the Lagrangians that produce terms that act as improvements.} whenever the matter Lagrangian is of the form
\beq
\mathcal{L}^{improv.}=\mathcal{L}_{M}(g_{\alpha\beta}, \Gamma^{\lambda}{}_{\mu\nu},\Phi)=\mathcal{L}_{M}(g_{\alpha\beta}, S_{\mu\nu}{}^{\lambda}, Q_{[\nu\mu]\alpha},\Phi)
\eeq
where $\Phi$ collectively denotes all matter fields that may be present. To illustrate the point that indeed such nonmetricity couplings act as improvement terms let us give an example. The simplest case is that of a scalar field coupled to the connection. Let us consider the coupling 
\beq
\mathcal{L}^{Q}=Q_{[\nu\mu]\alpha}g^{\alpha\mu}\partial^{\nu}\phi 
\eeq
We shall show that this is an improvement term by computing the double derivative of the associate hypermomentum tensor. Varying the above term with respect to the connection and consequently taking the flat space limit, we find 
\beq
\Delta_{\lambda}{}^{\mu\nu}=
-\frac{2}{\sqrt{-g}}\frac{\delta (\sqrt{-g}\mathcal{L}^{Q})}{\delta \Gamma^{\lambda}{}_{\mu\nu}}\Big|_{\Gamma=0,\; g=\eta}=-2\delta_{\lambda}^{\mu}\partial^{\nu}\phi+\delta_{\lambda}^{\nu}\partial^{\mu}\phi+\eta^{\mu\nu}\partial_{\lambda}\phi \label{hypQ}
\eeq
Then, given the commutativity of partial derivatives  we see that
\beq
\partial_{\mu}\partial_{\nu}\Delta_{\lambda}{}^{\mu\nu}=-2 \partial_{\lambda}\partial^{2}\phi+ \partial_{\lambda}\partial^{2}\phi +\partial^{2}\partial_{\lambda}\phi\equiv 0
\eeq
off-shell, proving our statement that it is indeed an improvement term! One could object that the resulting tensor is not antisymmetric in $\mu,\nu$. However, one should note that since only nonmetricity is involved, the hypermomentum $\Delta^{\alpha\mu\nu}$ must also be symmetric in $\alpha,\mu$, consequently, the resulting tensor has a mixed symmetry and it can be expressed as
\beq
\Delta^{\alpha\mu\nu}=-4 (\partial^{[\nu}\phi)\eta^{(\mu]\lambda)}
\eeq

Therefore, nonmetricity couplings can also produce improvement terms that do not alter the conservation law of the energy-momentum tensor. Nevertheless, one should be cautious when constructing such couplings with nonmetricity, whereas with torsion the situation is pretty straightforward. This places torsion on a  somewhat privileged, more straightforward, position when constructing  couplings producing improvement terms.

However, this nonmetricity coupling does have some quite important physical consequences. To be more specific, let us consider a free scalar field with the usual flat space action
\beq
S_{\phi}=-\frac{1}{2}\int d^{n}x (\partial_{\mu}\phi)({\partial^{\mu}\phi}) =-\frac{1}{2}\int d^{n}x (\partial \phi)^2 \label{examplephi}
\eeq
The affine lift leaves it essentially unaffected since only one partial derivative acts one $\phi$. Let us add to it the above coupling term multiplied also with $\phi$ to have proper dimensions. The uplifted action with the added coupling reads
\beq
S_{\phi}=\int d^{n}x \sqrt{-g} \Big[-\frac{1}{2} (\partial \phi)^{2}+\lambda 
Q_{[\nu\mu]\alpha}g^{\alpha\mu}\phi\partial^{\nu}\phi \Big]
\eeq
where $\lambda$ is a dimensional constant to be specified shortly. The associated hypermomentum in the flat space limit is computed to be
\beq
\Delta^{\nu\mu\lambda}=\lambda \Big( \eta^{\mu\lambda}\phi \partial^{\nu}\phi+\eta^{\nu\lambda}\phi \partial^{\mu}\phi-2 \eta^{\mu\nu}\phi \partial^{\lambda}\phi \Big)
\eeq
and the canonical energy-momentum tensor has the usual form\footnote{Note that in this case it just so happens to be symmetric by default.}
\beq
t^{\mu\nu}=\partial^{\mu}\phi\partial^{\nu}\phi-\frac{1}{2}\eta^{\mu\nu}(\partial \phi)^{2}
\eeq
Then substitution of the latter two equations into the master relation (\ref{flatCL}) yields the on-shell metrical energy-momentum tensor
\beq
T^{\mu\nu}=\partial^{\mu}\phi\partial^{\nu}\phi-\frac{1}{2}\eta^{\mu\nu}(\partial \phi)^{2}-\lambda \Big( \phi \partial^{\mu}\partial^{\nu}\phi
+\partial^{\mu}\phi \partial^{\nu}\phi-\eta^{\mu\nu}(\partial \phi)^{2}\Big) \label{Tconfcoupl}
\eeq
By demanding a vanishing trace $T=T^{\mu\nu}g_{\mu\nu}\equiv 0$ we find  $\lambda$ to be given by
\beq
\lambda=\frac{1}{2}\left(\frac{n-2}{n-1}\right)
\eeq
and consequently 
\beq
T^{\mu\nu}=\partial^{\mu}\phi\partial^{\nu}\phi-\frac{1}{2}\eta^{\mu\nu}(\partial \phi)^{2}-\frac{1}{2}\left(\frac{n-2}{n-1}\right)\Big( \phi \partial^{\mu}\partial^{\nu}\phi
+\partial^{\mu}\phi \partial^{\nu}\phi-\eta^{\mu\nu}(\partial \phi)^{2}\Big)
\eeq
Using also the identity $\partial^{\mu}\partial^{\nu}\phi^{2}=2(\partial^{\mu}\partial^{\nu}\phi+\phi \partial^{\mu}\partial^{\nu}\phi)$ and the on-shell relation $\partial^{2}\phi^{2}=2 (\partial \phi)^{2}$ the latter takes the more familiar form
\beq
T^{\mu\nu}=\partial^{\mu}\phi\partial^{\nu}\phi-\frac{1}{2}\eta^{\mu\nu}(\partial \phi)^{2}-\frac{1}{4}\left(\frac{n-2}{n-1}\right)\Big( \partial^{\mu}\partial^{\nu}\phi^{2}-\eta^{\mu\nu}\partial^{2}\phi^2\Big) \label{Tconf}
\eeq
of the traceless tensor of the conformally coupled scalar field \cite{CallanColemanJackiw1970}. The above stress energy-tensor is of course traceless, symmetric and given the improvement character of this hypermomentum, it is also conserved! Thus, the improvement couplings coming from nonmetricity gave a metrical energy-momentum tensor that is also conserved. This was not possible with the torsion coupling considered in \cite{Iosifidis:2025sjx} where, there, only the canonical energy-momentum tensor could be made traceless. The example here shows the importance of non-metric couplings for this achievement. But why did this prescription with this particular combination worked? Let us elaborate. Firstly,
it is important to emphasize that the usual Weyl covariant derivative \cite{Hecht1992,iosifidis2019scale}
\beq
D_{\mu}^{W}:=\partial_{\mu}+\frac{(2-n)}{4n}Q_{\mu} \label{Weylder}
\eeq
even thought it produces a traceless energy-momentum tensor $T=0$, it fails to preserve its conservation. The reason being that the associated hypermomentum that it produces is not of the improvement type, namely, its double derivative is not of-shell zero. Our proposed term here $Q_{[\nu\mu]\alpha}g^{\alpha\mu}\phi\partial^{\nu}\phi$ has a double success, not only does it produce a traceless tensor, but also preserves its conservation $\partial_{\mu}T^{\mu\nu}=0$. The last consequence comes, of course, by its very construction. As for the traceless-ness property, the reason is very simple. The interaction term can also be written as
\beq
\lambda 
Q_{[\nu\mu]\alpha}g^{\alpha\mu}\phi\partial^{\nu}\phi =\frac{1}{4}\left( \frac{n-2}{n-1}\right)(Q_{\mu}-q_{\mu})\phi \partial^{\mu}\phi
\eeq
which suggests the introduction of an \underline{'Improved Weyl Derivative'}\footnote{The name is self-explanatory; it produces a hypermomentum of improvement type and also guarantees conformal invariance of the total action. Note also that only for Weyl non-metricity $Q_{\alpha\mu\nu}=\frac{Q_{\alpha}}{n}g_{\mu\nu}$, this derivative reduces to (\ref{Weylder}). This also proves that Weyl geometries are inadequate to produce improvement terms since for such restricted geometries $Q_{[\alpha\mu]\nu}g^{\mu\nu}\equiv 0$ identically.}
\beq
\boxed{\bold{D}_{\mu}:=\partial_{\mu}-\frac{1}{4}\left( \frac{n-2}{n-1}\right) (Q_{\mu}-q_{\mu}) }\label{imprWeyl}
\eeq
Then the action 
\beq
S_{imp+conf}=-\frac{1}{2}\int d^{n}x \sqrt{-g}g^{\mu\nu}(\bold{D}_{\mu}\phi)(\bold{D}_{\nu}\phi)=-\frac{1}{2}\int d^{n}x \sqrt{-g}(\bold{D} \phi)^{2} \label{acimprconf}
\eeq
is conformally invariant and its associated hypermomentum is of the improved type.
Therefore, the significance of the derivative is twofold; it produces a hypermomentum of the improvement type and also defines a conformal invariant action, namely the above action (\ref{acimprconf}) is invariant under the conformal transformation
\beq
g_{\mu\nu}\rightarrow e^{2 \theta}g_{\mu\nu}
\eeq
which sends
\beq
Q_{\mu}\rightarrow Q_{\mu}-2 n\partial_{\mu}\theta\;\;, \;\; q_{\mu}\rightarrow q_{\mu}-2 \partial_{\mu}\theta
\eeq
accompanied by the field transformation $\phi \rightarrow e^{\frac{(2-n)}{2}\theta}\phi$. This example also illustrates the significance of nonmetricity in deriving a traceless and conserved energy-momentum tensor. We would like to stress here this last property, namely the capability of nonmetricity to produce improvement terms, something that was not realized in \cite{Iosifidis:2025sjx}. We also observe that the same term arises as the analogue of the conformally coupled scalar field term $\phi^{2}\tilde{R}$ in the symmetric teleparallel gravity (where both curvature and torsion are set to zero). Concluding, we have introduced a new kind of improved Weyl derivative, as given by (\ref{imprWeyl}), which guarantees a traceless, as well as conserved energy-momentum tensor for the free scalar field. For this construction, the existence of nonmetricity is essential. In principal, it would be possible to extend this to higher derivative scalar field theories.

%Indeed in  the Ricci scalar decomposition for vanishing torsion $R=\tilde{R}+Q+\tilde{\nabla}_{\mu}(q^{\mu}-Q_{\mu})$,
%for a flat ($R^{\alpha}{}_{\mu\nu\lambda}=0$) connection it follows that
%\beq
%\phi^{2}\tilde{R}=
%\eeq

\subsection{The relevant couplings in the matter Lagrangian expansion}

In order to connect with the previous subsection regarding the relevant connection couplings that produce improvement terms, let us consider a generic affinely uplifted material Lagrangian $\bold{\mathcal{L}^{aff}_{M}}(g_{\alpha\beta},\Gamma^{\lambda}{}_{\alpha\beta},\Phi)$. We  expand the latter according to
\beq
\bold{\mathcal{L}^{aff}_{\text{M}}}(g_{\alpha\beta},\Gamma^{\lambda}{}_{\alpha\beta},\Phi)=\mathcal{L}^{(0)}(g_{\alpha\beta},\Phi)+S_{\alpha\beta}{}^{\gamma}\Xi_{\gamma}{}^{\alpha\beta}+Q_{\rho\alpha\beta}Z^{\rho\alpha\beta}+\mathcal{L}^{(2)}(g_{\alpha\beta},\Gamma^{\lambda}{}_{\alpha\beta},\Phi) \label{expandedL}
\eeq
where we have identified the conjugates
\beq
\Xi_{\lambda}{}^{\mu\nu}\equiv \frac{\partial \bold{\mathcal{L}^{aff}_{\text{M}}} }{\partial S_{\mu\nu}{}^{\lambda}}\Big|_{\Gamma=0} \;\;, \;\;\; Z^{\lambda\mu\nu}\equiv \frac{\partial \bold{\mathcal{L}^{aff}_{\text{M}}} }{\partial Q_{\lambda\mu\nu}}\Big|_{\Gamma=0}
\eeq
evaluated around Riemannian space, 
with the symmetry properties $\Xi_{\lambda}{}^{\mu\nu}=\Xi_{\lambda}{}^{[\mu\nu]}$, $Z_{\lambda}{}^{\mu\nu}=Z_{\lambda}{}^{(\mu\nu)}$ by construction. In (\ref{expandedL}), the last term
 $\mathcal{L}^{(2)}$ is the part of the Lagrangian containing second order and higher in torsion and nonmetricity. Since at the end of the day we will take the flat space limit, these higher order terms are totally irrelevant and do not contribute to the form of hypermomentum. What is essential, is the linear part in the connection which is captured in the second and third parts of (\ref{expandedL}). The variation of this linear coupling is independent of the connection and is the only one that contributes in the flat space Minkowski limit. The hypermomentum for the above Lagrangian, after taking the flat space limit,  reads
\beq
\Delta_{\lambda}{}^{\mu\nu}=-2\Xi_{\lambda}{}^{\mu\nu}-4 Z^{\nu\mu}{}{}_{\lambda} \label{hyperXZ}
\eeq
Again, in order for the latter to be an improvement term, it needs to have an identically vanishing double derivative over the last two indices. Given the antisymmetry of $\Xi$ in the last pair of indices, we have $\partial_{\mu}\partial_{\nu}\Xi_{\lambda}{}^{\mu\nu}\equiv 0$ identically, so the condition we need to impose translates to
\beq
\partial_{\mu}\partial_{\nu}Z^{\mu\nu}{}{}_{\lambda}=0
\eeq
Note the index placing. The latter condition is certainly true\footnote{Let us remark again that this is not the most general solution, in other words the antisymmetric in the first two indices  is  a capable but not a necessary condition.} for tensors with the property
\beq
Z^{\mu\nu}{}{}_{\lambda}=Z^{[\mu\nu]}{}{}_{\lambda}
\eeq
namely  when the Z tensor is antisymmetric in its first indices, the full Lagrangian (\ref{expandedL}) produces a hypermomentum that acts as an improvement term, symmetrizing the canonical energy-momentum tensor. To connect to the nonmetricity coupling of the previous subsection, let us again consider the case of a scalar field. Since $Z^{\mu\nu\lambda}$ is independent of the connection, it must be built entirely from the metric and the derivative of the scalar field. Furthermore, since it must be symmetric in $\nu,\lambda$ and antisymmetric in $\mu,\nu$, therefore, up to a numerical factor, it is fixed to be of the form $Z^{\mu\nu\lambda}\propto (\partial^{[\mu}\phi)\eta^{(\nu]\lambda)}\propto -2\delta_{\lambda}^{\mu}\partial^{\nu}\phi+\delta_{\lambda}^{\nu}\partial^{\mu}\phi+\eta^{\mu\nu}\partial_{\lambda}\phi$,  in perfect agreement with the result (\ref{hypQ}).

The expression (\ref{hyperXZ}) also tells us something very significant; substituting this form into (\ref{cl2}), and taking the antisymmetric part of the latter, we obtain for the antisymmetric part of the canonical tensor
\beq
t^{[\mu\nu]}=-\partial_{\lambda}\Xi^{[\nu\mu]\lambda}-2\partial_{\lambda}Z^{\lambda[\nu\mu]}
\eeq
but recall that $Z^{\lambda\mu\nu}$ is symmetric in $\mu,\nu$ and therefore the second term on the right-hand side vanishes identically, so we are left with
\beq
t^{[\mu\nu]}=\partial_{\lambda}\Xi^{[\mu\nu]\lambda}=\frac{1}{2}\partial_{\lambda}\sigma^{\nu\mu\lambda}
\eeq
namely only the torsion coupling gives rise to an antisymmetric part for the canonical energy-momentum tensor. This is in perfect agreement with the well-known fact that only torsion couples to spin for minimally coupled theories.

\subsection{An alternative direct route to the symmetrized tensor}

Looking at the master equation (\ref{flatCL}a) one might get the impression that the symmetric energy-momentum tensor $T^{\mu\nu}$ that we are after can only indirectly derived by the knowledge of the canonical $t^{\mu\nu}$ and the hypermomentum $\Delta_{\lambda}{}^{\mu\nu}$. This is not the case. One also has direct access to the metrical tensor. To see this consider a matter Lagrangian $\mathcal{L}_{M}=\mathcal{L}_{M}(g_{\mu\nu},S_{\alpha\beta}{}{}^{\gamma},Q_{\kappa}{}^{\lambda\rho},\Phi)$. Note that the dependence on the partial derivatives of the metric can only come through the covariant form of nonmetricity and therefore, using the chain rule, the metric variation is computed as
\beq
\frac{\delta (\sqrt{-g}\mathcal{L}_{M})}{\delta g^{\mu\nu}}=-\frac{\sqrt{-g}}{2}\mathcal{L}_{M}g_{\mu\nu}+\sqrt{-g}\frac{\partial \mathcal{L}_{M}}{\partial g^{\mu\nu}}+\sqrt{-g}\Big( 2 S_{\alpha}+\frac{Q_{\alpha}}{2}-\nabla_{\alpha}\Big) \frac{\partial \mathcal{L}_{M}}{\partial Q_{\alpha}{}^{\mu\nu}}+s.t.
\eeq
where we should partial integration and ignored a surface term (denoted as $s.t.$). As a result, given the definition (\ref{metrical}), the metrical energy-momentum tensor has the flat space expression
\beq
\boxed{T_{\mu\nu}=g_{\mu\nu}\mathcal{L}_{M}-2 \frac{\partial \mathcal{L}_{M}}{\partial g^{\mu\nu}}+ 2\partial_{\alpha}q^{\alpha}{}_{\mu\nu} } \label{Tmndirect}
\eeq
where we have defined
\beq
q^{\alpha}{}_{\mu\nu}:=\left(\frac{\partial \mathcal{L}_{M}}{\partial Q_{\alpha}{}^{\mu\nu}}\right)_{g=\eta\;, \Gamma=0}
\eeq
as the nonmetric remnant of the affine lift. For instance, if we apply this to our example (\ref{examplephi}) we compute, for this case,
\beq
\frac{\partial \mathcal{L}_{M}}{\partial Q_{\alpha}{}^{\mu\nu}}=\frac{1}{2}g_{\mu\nu}\partial^{\alpha}\phi -\frac{1}{4}\delta^{\alpha}_{\nu}\partial_{\mu}\phi
-\frac{1}{4}\delta^{\alpha}_{\mu}\partial_{\nu}\phi
\eeq
Using this along with trivial computation of partial derivatives, the relation (\ref{Tmndirect}) gives us
\beq
T^{\mu\nu}=\partial^{\mu}\phi\partial^{\nu}\phi-\frac{1}{2}\eta^{\mu\nu}(\partial \phi)^{2}-\lambda \Big( \phi \partial^{\mu}\partial^{\nu}\phi
+\partial^{\mu}\phi \partial^{\nu}\phi-\eta^{\mu\nu}(\partial \phi)^{2}\Big)
\eeq
which is identical to the expression (\ref{Tconfcoupl}). We see therefore how using the formula (\ref{Tmndirect}) gives us direct access to the improved symmetric tensor. Further requirement of vanishing trace gives us the expression (\ref{Tconf}). One may then wonder why do we have to bother by hypermonentum in the first place since (\ref{Tmndirect}) gives us the symmetric energy-momentum tensor directly. The simple answer is that the hypermomentum has to be of the improved type in order for the resulting symmetric tensor to be conserved. If the hypermomentum is arbitrary and not of the improvement type, this tensor is still going to be symmetric but non-conserved. Therefore it is essential to have the proper hypermomentum form in order for the conservation to continue to hold true.

Let us reflect upon the master equation
\beq
t^{\mu\nu}=T^{\mu\nu}+\frac{1} {2}\partial_{\alpha}\Delta^{\nu\mu\alpha} 
\eeq
We must emphasize that in this flat space expression the canonical tensor is prescribed and is the same regardless of the affine couplings considered for the affine lift. It is insensitive to the symmetrization process. Under an affine lift of the geometry, both $T^{\mu\nu}$ and $\Delta_{\lambda}{}^{\mu\nu}$ change but they do so in such a way so as to ensure that $t^{\mu\nu}$ remains unaffected. Our example above clarifies this subtle point. As a matter of fact, we see that $t^{\mu\nu}$ remains unaffected under the simultaneous replacements
\beq
T^{\mu\nu}\rightarrow T^{\mu\nu}-\frac{1}{2}\partial_{\alpha}\chi^{\nu\mu\alpha}\;\;, \;\; \Delta^{\nu\mu\alpha} \rightarrow \Delta^{\nu\mu\alpha} +\chi^{\nu\mu\alpha}
\eeq
This is actually a re-localization of energy with the above transformations being another manifestation of the  so-called pseudo-gauge transformations \cite{Hehl1976EnergyTensor}.\footnote{The term pseudo-gauge transformation is broadly used in spin hydrodynamics, but the more appropriate term is the 're-localization' of energy. Indeed, this transformation rather relocates the
energy-momentum and the spin densities differently.}

%Another example worth mentioning is that of a Yang-Mils Lagrangian

\subsection{The step by step procedure}

Let us gather here the basic steps one has to follow in order to derive the improved symmetric energy-moemntum tensor of any given field theory in Minkowski space. 

%\textbf{Recap of the algorithm}

\begin{enumerate}
    \item

     \textbf{Step 1.} Perform an \textit{affine lift} (i.e. 'Go Metric-Affine'), that is, simultaneously promote
  $
\eta_{\mu\nu} \rightarrow g_{\mu\nu}
  $
and 
$
\partial_{\mu}\rightarrow \nabla_{\mu}
$
where $\nabla_{\mu}$ is a generic torsionful and non-metric covariant derivative. Write down the affinely-lifted  matter action $\Rightarrow$ $S_{M}[g,\Gamma,\phi^{A}]$.

\item  \textbf{Step 2.} Vary the  resulting action wrt affine connection to obtain the hypermomentum and consequently go flat, i.e. set $g_{\mu\nu}\equiv \eta_{\mu\nu}$ and $\Gamma^{\lambda}{}_{\mu\nu}\equiv 0$ after the variation has been performed, viz. 
\beq
\Delta_{\lambda}{}^{\mu\nu}:=-\frac{2}{\sqrt{-g}}\frac{\delta  S_{M}}{\delta \Gamma^{\lambda}{}_{\mu\nu}} \Big|_{g=\eta, \Gamma=0}
\eeq

\item  \textbf{Step 3.} Substitute the form of the hypermomentum found in the previous step into the flat space conservation law (\ref{cl2}) and solve for $T_{\mu\nu}$.

 \item \textbf{Outcome:} The resulting tensor $T_{\mu\nu}$ is the desired symmetric energy-momentum tensor which will be also conserved provided the hypermomentum has the of-shell property $\partial_{\mu}\partial_{\nu}\Delta^{\alpha\mu\nu}\equiv 0$ which is guaranteed if $\Delta_{\alpha\mu\nu}=\Delta_{\alpha[\mu\nu]}$ or if the couplings involve only the torsion tensor and the $Q_{[\alpha\beta]\gamma}$ part of nonmetricity. 

    \end{enumerate}

\textit{Corollary}. The Belinfante-Rosenfeld realtion is the flat-spacetime manifestation of the metric-affine hypermomentum conservation law (\ref{cc2}). When the explicit connection-matter couplings involve the torsion alone, or the $Q_{[\alpha\beta]\gamma}$ part of nonmetricity, then the symmetric tensor is also conserved, and hypermomentum justifies as the proper improvement term.

\section{conclusions}

In this work, we have shown that the Belinfante--Rosenfeld symmetrization procedure can be understood as the flat-spacetime limit of a metric-affine conservation law obtained through an affine lift of flat-spacetime field theories. By treating the affine connection as an independent geometric variable and implementing minimal coupling, variation of the action with respect to the connection naturally gives rise to the hypermomentum current. The corresponding conservation identity then reduces, in the flat-spacetime limit, to the Belinfante--Rosenfeld relation, with the divergence of the hypermomentum current reproducing the familiar Belinfante improvement term.

Our construction provides a geometric origin for the Belinfante improvement, which is traditionally introduced to obtain a symmetric and conserved energy--momentum tensor suitable for coupling to gravity \cite{Belinfante1940,Belinfante1940b,Rosenfeld1940}. Rather than arising as an \textit{ad hoc} modification of the canonical tensor, the improvement emerges directly from the response of matter to an independent affine connection. In this way, the hypermomentum current acquires a transparent physical interpretation as the quantity whose conservation properties encode the symmetrization of the energy--momentum tensor.

The results also elucidate the relationship between the canonical currents of flat-spacetime field theory and the hypermomentum current that arises when the theory is lifted to a metric-affine spacetime. In particular, the Belinfante improvement is recovered as the flat-spacetime remnant of the metric-affine conservation law. Therefore, the Belinfante--Rosenfeld construction is naturally embedded within the broader framework of metric-affine gauge theories of gravity.
The explicit examples of QED and QCD demonstrate that the formalism reproduces the standard Belinfante-improved energy--momentum tensor for gauge theories while simultaneously identifying the corresponding hypermomentum current. This suggests that the affine-lift construction is sufficiently general to accommodate the field theories of the Standard Model and provides a unified geometric perspective on their conserved currents. Furthermore, we explicitly showed that non-metricity couplings can also create improvement terms. In particular, considering the free scalar field, we proved how, through a generalized improved Weyl derivative, as given by (\ref{imprWeyl}), one can arrive at an energy-momentum tensor that is symmetric, conserved, and traceless. This achievement, would not be possible through torsion couplings; the introduction of nonmetricity is essential in this setting. 

Several extensions of the present work deserve further investigation. It would be interesting to examine theories with non-minimal couplings or higher-derivative interactions, where additional contributions to the hypermomentum may arise, as well as theories containing higher-spin fields. It would also be worthwhile to explore whether the affine-lift approach can shed new light on quantum aspects of conserved currents, including trace and gravitational anomalies as in \cite{Miyashita2025TraceAnomaly,Bahamonde2025TraceAnomaly}, and on the role of hypermomentum in generalized gravitational theories.

Overall, the results presented here establish that the Belinfante--Rosenfeld symmetrization procedure is not merely a convenient field-theoretic prescription but the  flat-spacetime manifestation of a more fundamental conservation law associated with the generalized geometry. This connection provides a unified geometric understanding of the relationship between canonical energy-momentum, hypermomentum, and the gravitational coupling of matter, thereby further strengthening the connection between relativistic field theory and metric-affine gravity.

\section{Acknowledgments}
 This work was supported by the  Istituto Nazionale di Fisica Nucleare (INFN), Sezioni  di Napoli,  {\it Iniziative Specifiche} SKY.  I would like to thank Friedrich Hehl and also  Kostas Siampos and Anastasios Petkou for useful discussions.

\bibliographystyle{unsrt}
\bibliography{ref}

@article{Forger2004Currents,
  author       = {Michael Forger and Hartmann R\"omer},
  title        = {Currents and the Energy-Momentum Tensor in Classical Field Theory: A Fresh Look at an Old Problem},
  journal      = {Annals of Physics},
  volume       = {309},
  number       = {2},
  pages        = {306--389},
  year         = {2004},
  doi          = {10.1016/j.aop.2003.08.011},
  eprint       = {hep-th/0307199},
  archivePrefix= {arXiv}
}

@incollection{GotayMarsden1992,
  author    = {Mark J. Gotay and Jerrold E. Marsden},
  title     = {Stress-Energy-Momentum Tensors and the Belinfante-Rosenfeld Formula},
  booktitle = {Mathematical Aspects of Classical Field Theory},
  series     = {Contemporary Mathematics},
  volume     = {132},
  pages      = {367--392},
  year       = {1992},
  publisher  = {American Mathematical Society},
  address    = {Providence, RI},
  editor     = {Mark J. Gotay and Jerrold E. Marsden and Vincent Moncrief},
  doi        = {10.1090/conm/132/1188448}
}

@article{Hecht1992,
  author  = {Ralf Hecht and Friedrich W. Hehl and J. Dermott McCrea and Eckehard W. Mielke and Yuval Ne'eman},
  title   = {Improved Energy-Momentum Currents in Metric-Affine Spacetime},
  journal = {Physics Letters A},
  volume  = {172},
  number  = {1--2},
  pages   = {13--20},
  year    = {1992},
  doi     = {10.1016/0375-9601(92)90182-L}
}

@article{Julia:1998ys,
    author = "Julia, B. and Silva, S.",
    title = "{Currents and superpotentials in classical gauge invariant theories. 1. Local results with applications to perfect fluids and general relativity}",
    eprint = "gr-qc/9804029",
    archivePrefix = "arXiv",
    reportNumber = "LPTENS-98-06",
    doi = "10.1088/0264-9381/15/8/006",
    journal = "Class. Quant. Grav.",
    volume = "15",
    pages = "2173--2215",
    year = "1998"
}

@article{Bahamonde2025TraceAnomaly,
  author  = {Sebastian Bahamonde and Yuichi Miyashita and Masahide Yamaguchi},
  title   = {Trace Anomaly in Metric-Affine Gravity},
  journal = {Physical Review D},
  volume  = {111},
  number  = {4},
  pages   = {044065},
  year    = {2025},
  month   = {February},
  doi     = {10.1103/PhysRevD.111.044065},
  eprint  = {2409.05499},
  archivePrefix = {arXiv},
  primaryClass = {gr-qc}
}

@phdthesis{Miyashita2025TraceAnomaly,
  author       = {Yuichi Miyashita},
  title        = {Trace Anomaly in Metric-Affine Gravity},
  school       = {Institute of Science Tokyo},
  year         = {2025},
  month        = {February},
  type         = {PhD thesis}
}

@article{GamboaSaravi2004,
  author       = {R. E. Gamboa Sarav\'i},
  title        = {On the Energy-Momentum Tensor},
  journal      = {Journal of Physics A: Mathematical and General},
  volume       = {38},
  number       = {4},
  pages        = {919--930},
  year         = {2005},
  doi          = {10.1088/0305-4470/38/4/017},
  eprint       = {math-ph/0306020},
  archivePrefix= {arXiv}
}

@article{Pons2009Noether,
  author       = {Josep M. Pons},
  title        = {Noether Symmetries, Energy-Momentum Tensors and Conformal Invariance in Classical Field Theory},
  journal      = {Journal of Physics A: Mathematical and Theoretical},
  volume       = {43},
  number       = {10},
  pages        = {105003},
  year         = {2010},
  doi          = {10.1088/1751-8113/43/10/105003},
  eprint       = {0902.4871},
  archivePrefix= {arXiv},
  primaryClass = {hep-th}
}

@article{HEHL1977432,
title = {Hadron dilation, shear and spin as components of the intrinsic hypermomentum current and metric-affine theory of gravitation},
journal = {Physics Letters B},
volume = {71},
number = {2},
pages = {432-434},
year = {1977},
issn = {0370-2693},
doi = {https://doi.org/10.1016/0370-2693(77)90260-X},
url = {https://www.sciencedirect.com/science/article/pii/037026937790260X},
author = {F.W. Hehl and E.A. Lord and Y. Ne'eman}
}

@article{Nielsen1977,
  author  = {N. K. Nielsen},
  title   = {The Energy-Momentum Tensor in a Non-Abelian Quark Gluon Theory},
  journal = {Nuclear Physics B},
  volume  = {120},
  number  = {2},
  pages   = {212--220},
  year    = {1977},
  doi     = {10.1016/0550-3213(77)90040-2}
}

@article{Lorce:2021xku,
    author = "Lorc{\'e}, C{\'e}dric and Metz, Andreas and Pasquini, Barbara and Rodini, Simone",
    title = "{Energy-momentum tensor in QCD: nucleon mass decomposition and mechanical equilibrium}",
    eprint = "2109.11785",
    archivePrefix = "arXiv",
    primaryClass = "hep-ph",
    doi = "10.1007/JHEP11(2021)121",
    journal = "JHEP",
    volume = "11",
    pages = "121",
    year = "2021"
}

@article{Becattini:2018duy,
    author = "Becattini, F. and Florkowski, Wojciech and Speranza, Enrico",
    title = "{Spin tensor and its role in non-equilibrium thermodynamics}",
    eprint = "1807.10994",
    archivePrefix = "arXiv",
    primaryClass = "hep-th",
    doi = "10.1016/j.physletb.2018.12.016",
    journal = "Phys. Lett. B",
    volume = "789",
    pages = "419--425",
    year = "2019"
}

@article{Embacher1986,
  author  = {Embacher, Hans Georg and Gr{\"u}bl, Gebhard and Patek, Rainer},
  title   = {Gauge-invariant energy-momentum tensor for massive QED},
  journal = {Physical Review D},
  volume  = {33},
  number  = {4},
  pages   = {1162--1165},
  year    = {1986},
  doi     = {10.1103/PhysRevD.33.1162}
}

@article{Freese:2025glz,
    author = "Freese, Adam",
    title = "{Reflections on Noether{\textquoteright}s second theorem and the energy-momentum tensor}",
    eprint = "2506.04510",
    archivePrefix = "arXiv",
    primaryClass = "hep-ph",
    reportNumber = "JLAB-THY-25-4355",
    doi = "10.1103/wmz3-lrbz",
    journal = "Phys. Rev. D",
    volume = "113",
    number = "1",
    pages = "016011",
    year = "2026"
}

@article{KugoOjima1979,
  author  = {Kugo, Taichiro and Ojima, Izumi},
  title   = {Local Covariant Operator Formalism of Non-Abelian Gauge Theories and Quark Confinement Problem},
  journal = {Progress of Theoretical Physics Supplement},
  volume  = {66},
  pages   = {1--130},
  year    = {1979},
  doi     = {10.1143/PTPS.66.1}
}

@article{Leader:2013jra,
    author = "Leader, E. and Lorc{\'e}, C.",
    title = "{The angular momentum controversy: What{\textquoteright}s it all about and does it matter?}",
    eprint = "1309.4235",
    archivePrefix = "arXiv",
    primaryClass = "hep-ph",
    doi = "10.1016/j.physrep.2014.02.010",
    journal = "Phys. Rept.",
    volume = "541",
    number = "3",
    pages = "163--248",
    year = "2014"
}

@article{Montesinos:2006th,
    author = "Montesinos, Merced and Flores, Ernesto",
    title = "{Symmetric energy-momentum tensor in Maxwell, Yang-Mills, and proca theories obtained using only Noether's theorem}",
    eprint = "hep-th/0602190",
    archivePrefix = "arXiv",
    journal = "Rev. Mex. Fis.",
    volume = "52",
    pages = "29--36",
    year = "2006"
}

@article{Ji:1996ek,
    author = "Ji, Xiang-Dong",
    title = "{Gauge-Invariant Decomposition of Nucleon Spin}",
    eprint = "hep-ph/9603249",
    archivePrefix = "arXiv",
    reportNumber = "MIT-CTP-2517",
    doi = "10.1103/PhysRevLett.78.610",
    journal = "Phys. Rev. Lett.",
    volume = "78",
    pages = "610--613",
    year = "1997"
}

@article{Zoller:2014dca,
    author = "Zoller, Max F.",
    title = "{OPE of the energy-momentum tensor correlator and the gluon condensate operator in massless QCD to three-loop order}",
    eprint = "1407.6921",
    archivePrefix = "arXiv",
    primaryClass = "hep-ph",
    reportNumber = "TTP14-023, SFB-CPP-14-56",
    doi = "10.1007/JHEP10(2014)169",
    journal = "JHEP",
    volume = "10",
    pages = "169",
    year = "2014"
}

@article{Iosifidis:2025sjx,
    author = "Iosifidis, Damianos and Karydas, Manthos and Petkou, Anastasios and Siampos, Konstantinos",
    title = "{Geometric origin of the energy-momentum tensor improvement terms}",
    eprint = "2503.21609",
    archivePrefix = "arXiv",
    primaryClass = "hep-th",
    doi = "10.1103/wby2-d33f",
    journal = "Phys. Rev. D",
    volume = "112",
    number = "2",
    pages = "025007",
    year = "2025"
}

@article{Blaschke2016,
  author = {Daniel N. Blaschke and Fran\c{c}ois Gieres and Meril Reboud and Manfred Schweda},
  title = {The Energy-Momentum Tensor(s) in Classical Gauge Theories},
  journal = {Nucl. Phys. B},
  volume = {912},
  pages = {192--223},
  year = {2016},
  doi = {10.1016/j.nuclphysb.2016.07.001}
}

@article{Noether1918,
  author  = {Emmy Noether},
  title   = {Invariante Variationsprobleme},
  journal = {Nachrichten von der Gesellschaft der Wissenschaften zu Göttingen, Mathematisch-Physikalische Klasse},
  pages   = {235--257},
  year    = {1918}
}

@book{Weinberg1995,
  author    = {Steven Weinberg},
  title     = {The Quantum Theory of Fields, Volume I: Foundations},
  publisher = {Cambridge University Press},
  year      = {1995}
}

@book{PeskinSchroeder1995,
  author    = {Michael E. Peskin and Daniel V. Schroeder},
  title     = {An Introduction to Quantum Field Theory},
  publisher = {Addison-Wesley},
  year      = {1995}
}

@article{Belinfante1940,
  author  = {F. J. Belinfante},
  title   = {On the Current and the Density of the Electric Charge, the Energy, the Linear Momentum and the Angular Momentum of Arbitrary Fields},
  journal = {Physica},
  volume  = {7},
  number  = {5},
  pages   = {449--474},
  year    = {1940},
  doi     = {10.1016/S0031-8914(40)90091-X}
}

@article{Montenegro:2026phf,
    author = "Montenegro, David and Savioli, Mariana Julia Pereira Dos Dores and Torrieri, Giorgio",
    title = "{Gaussian pseudogauge invariant hydrodynamics with spin}",
    eprint = "2604.07657",
    archivePrefix = "arXiv",
    primaryClass = "hep-th",
    month = "4",
    year = "2026"
}

@article{Zhang:2026DilationSpin,
  author        = {Zhong-Hua Zhang and Xi-Hu Lv and Xu-Guang Huang},
  title         = {Hydrodynamics of Dilation and Spin Currents},
  eprint        = {2603.17794},
  archivePrefix = {arXiv},
  primaryClass  = {hep-th},
  year          = {2026},
  month         = mar,
  doi           = {10.48550/arXiv.2603.17794}
}

@article{Becattini:2025oyi,
    author = "Becattini, Francesco and Singh, Rajeev",
    title = "{On the local thermodynamic relations in relativistic spin hydrodynamics}",
    eprint = "2506.20681",
    archivePrefix = "arXiv",
    primaryClass = "nucl-th",
    doi = "10.1140/epjc/s10052-025-15071-3",
    journal = "Eur. Phys. J. C",
    volume = "85",
    number = "11",
    pages = "1338",
    year = "2025"
}

@incollection{Hehl:1990SpacetimeMicrostructure,
  author    = {Friedrich W. Hehl and Yuval Ne'eman},
  title     = {Spacetime as a Medium with Microstructure and Cosmic Defects},
  booktitle = {Analogies Between Condensed Matter and Cosmology},
  editor    = {M. Novello},
  publisher = {World Scientific},
  address   = {Singapore},
  year      = {1990},
  pages     = {429--448}
}

@article{Gronwald:1991st,
  author        = {Frank Gronwald and Friedrich W. Hehl},
  title         = {Stress and hyperstress as fundamental concepts in continuum mechanics and in relativistic field theory},
  journal       = {International Journal of Theoretical Physics},
  volume        = {35},
  number        = {12},
  pages         = {2617--2640},
  year          = {1996},
  doi           = {10.1007/BF02302377},
  eprint        = {gr-qc/9701054},
  archivePrefix = {arXiv},
  primaryClass  = {gr-qc}
}

@article{Belinfante1940b,
  author  = {F. J. Belinfante},
  title   = {On the Spin Angular Momentum of Mesons},
  journal = {Physica},
  volume  = {7},
  number  = {9},
  pages   = {887--898},
  year    = {1940},
  doi     = {10.1016/S0031-8914(40)90030-1}
}

@article{Rosenfeld1940,
  author  = {L{\'e}on Rosenfeld},
  title   = {Sur le tenseur d'impulsion-{\'e}nergie},
  journal = {M{\'e}moires de l'Acad{\'e}mie Royale des Sciences, des Lettres et des Beaux-Arts de Belgique},
  volume  = {18},
  pages   = {1--30},
  year    = {1940}
}

@article{Hehl1976,
  author  = {Friedrich W. Hehl and Paul von der Heyde and G. David Kerlick and James M. Nester},
  title   = {General Relativity with Spin and Torsion: Foundations and Prospects},
  journal = {Reviews of Modern Physics},
  volume  = {48},
  number  = {3},
  pages   = {393--416},
  year    = {1976},
  doi     = {10.1103/RevModPhys.48.393}
}

@article{Hehl1995,
  author  = {Friedrich W. Hehl and J. Dermott McCrea and Eckehard W. Mielke and Yuval Ne'eman},
  title   = {Metric-affine Gauge Theory of Gravity: Field Equations, Noether Identities, World Spinors, and Breaking of Dilation Invariance},
  journal = {Physics Reports},
  volume  = {258},
  number  = {1--2},
  pages   = {1--171},
  year    = {1995},
  doi     = {10.1016/0370-1573(94)00111-F}
}

@book{HehlObukhov2003,
  author    = {Friedrich W. Hehl and Yuri N. Obukhov},
  title     = {Foundations of Classical Electrodynamics: Charge, Flux, and Metric},
  publisher = {Birkh{\"a}user},
  address   = {Boston},
  year      = {2003},
  doi       = {10.1007/978-1-4612-0051-2}
}

@article{ObukhovRubilar2006,
  author  = {Yuri N. Obukhov and Guillermo F. Rubilar},
  title   = {Invariant conserved currents in gravity theories with local Lorentz and diffeomorphism symmetry},
  journal = {Physical Review D},
  volume  = {74},
  pages   = {064002},
  year    = {2006},
  doi     = {10.1103/PhysRevD.74.064002}
}

@article{hehl1995metric,
  title={Metric-affine gauge theory of gravity: field equations, Noether identities, world spinors, and breaking of dilation invariance},
  author={Hehl, Friedrich W and McCrea, J Dermott and Mielke, Eckehard W and Ne'eman, Yuval},
  journal={Physics Reports},
  volume={258},
  number={1-2},
  pages={1--171},
  year={1995},
  publisher={Elsevier}
}

@article{iosifidis2019metric,
  title={Metric-Affine Gravity and Cosmology/Aspects of Torsion and non-Metricity in Gravity Theories},
  author={Iosifidis, Damianos},
  journal={arXiv preprint arXiv:1902.09643},
  year={2019}
}

@article{CallanColemanJackiw1970,
  author  = {Callan, Curtis G. and Coleman, Sidney and Jackiw, Roman},
  title   = {A New Improved Energy-Momentum Tensor},
  journal = {Annals of Physics},
  volume  = {59},
  pages   = {42--73},
  year    = {1970},
  doi     = {10.1016/0003-4916(70)90394-5}
}

@article{Hehl1976EnergyTensor,
  author  = {Hehl, Friedrich W.},
  title   = {On the Energy Tensor of Spinning Massive Matter in Classical Field Theory and General Relativity},
  journal = {Reports on Mathematical Physics},
  volume  = {9},
  number  = {1},
  pages   = {55--82},
  year    = {1976},
  doi     = {10.1016/0034-4877(76)90016-1}
}

@article{hehl1976hypermomentum,
  title={On hypermomentum in general relativity I. The notion of hypermomentum},
  author={Hehl, Friedrich W and Kerlick, G David and von der Heyde, Paul},
  journal={Zeitschrift fuer Naturforschung A},
  volume={31},
  number={2},
  pages={111--114},
  year={1976},
  publisher={Verlag der Zeitschrift f{\"u}r Naturforschung}
}

@book{landau1975classical,
  author    = {L. D. Landau and E. M. Lifshitz},
  title     = {The Classical Theory of Fields},
  series    = {Course of Theoretical Physics},
  volume    = {2},
  edition   = {4},
  publisher = {Butterworth-Heinemann},
  address   = {Oxford},
  year      = {1975},
  isbn      = {9780750627689}
}

@article{iosifidis2019scale,
  title={Scale transformations in metric-affine geometry},
  author={Iosifidis, Damianos and Koivisto, Tomi},
  journal={Universe},
  volume={5},
  number={3},
  pages={82},
  year={2019},
  publisher={Multidisciplinary Digital Publishing Institute}
}

@article{obukhov2013conservation,
  title={Conservation laws in gravitational theories with general nonminimal coupling},
  author={Obukhov, Yuri N and Puetzfeld, Dirk},
  journal={Physical Review D},
  volume={87},
  number={8},
  pages={081502},
  year={2013},
  publisher={APS}
}

@article{hehl1978hypermomentum,
  title={Hypermomentum in hadron dynamics and in gravitation},
  author={Hehl, FW and Lord, EA and Ne'Eman, Y},
  journal={Physical Review D},
  volume={17},
  number={2},
  pages={428},
  year={1978},
  publisher={APS}
}

@article{Iosifidis:2020gth,
    author = "Iosifidis, Damianos",
    title = "{Cosmological Hyperfluids, Torsion and Non-metricity}",
    eprint = "2003.07384",
    archivePrefix = "arXiv",
    primaryClass = "gr-qc",
    doi = "10.1140/epjc/s10052-020-08634-z",
    journal = "Eur. Phys. J. C",
    volume = "80",
    number = "11",
    pages = "1042",
    year = "2020"
}

\end{document}